Title: In vacancies in InN grown by plasma-assisted molecular beam epitaxy

Authors: Floris Reurings[1], Filip Tuomisto[1], Chad S. Gallinat[2], Gregor Koblmüller[2,3], and James S. Speck[2]

[1]Department of Applied Physics, Aalto University, Finland

[2]Materials Department, University of California, Santa Barbara, USA

[3]Walter Schottky Institut and Physics Department, Technische Universität München, Germany

The authors have applied positron annihilation spectroscopy to study the effect of different growth conditions on vacancy formation in In- and N-polar InN grown by plasma-assisted molecular beam epitaxy. The results suggest that the structural quality of the material and limited diffusion of surface adatoms during growth dictate the In vacancy formation in low electron-density undoped epitaxial InN, while growth conditions and thermodynamics have a less important role, contrary to what is observed in, e.g., GaN. Further, the results imply that in high quality InN, the electron mobility is likely limited not by ionized point defect scattering, but rather by threading dislocations.

The growth of high quality InN is a somewhat challenging task. Bulk material is not available, but high quality films on sapphire (either directly or using various buffer layers) have been fabricated by molecular beam epitaxy (MBE) [1–4] and also by metal-organic chemical vapor deposition (MOCVD) [5, 6]. The properties of the material are greatly affected by the layer thickness and substrate material, as well as the growth temperature and stoichiometry. In this work, we apply positron annihilation spectroscopy to study the effect of different growth conditions, i.e., growth temperature, polarity and stoichiometry, film thickness, and substrate material on vacancy formation in InN grown by MBE. Indium vacancies and vacancy clusters have been identified in previous positron studies on both MBE and MOCVD grown InN [7–9]. High In vacancy concentrations have been observed to coexist with high free electron densities and low electron mobilities in MBE-InN [7], while their formation has been found to be independent of growth stoichiometry, but dependent on growth temperature, in MOCVD-InN [8].

Positron annihilation spectroscopy is an effective method for the investigation of vacancy-type defects in semiconductors [10]. Positrons implanted in a sample can get trapped and localized at neutral and negative vacancies due to the missing positive ion core. This results in observable changes in the measurable annihilation characteristics, i.e., the positron lifetime and the momentum distribution of the annihilating positron–electron pair. The annihilation data can be used to determine the vacancy concentration as well as to distinguish between different types of vacancies and their chemical environments.

In this letter, we show that the in-grown In vacancy concentrations [$V_{In}$] even in high quality (low carrier density, high mobility) MBE-InN are much higher than could be expected from the predicted formation energy and that there is no clear correlation between [$V_{In}$] and growth conditions, impurity content, or dislocation densities. Hence, in high quality InN, the electron mobility is likely limited not by ionized point defect scattering. As the formation of $V_{In}$ is not dictated by thermodynamic equilibrium concentrations, even considering the effects of dislocations, impurities of chemical potential given by the growth environment on the formation energy, we suggest a formation mechanism at the growth surface.

The In-polar InN layers studied in this work were grown by plasma-assisted molecular beam epitaxy (PA-MBE) either directly on a semi-insulating MOCVD-GaN template or using a MBE-GaN buffer layer, at temperatures in the range from 430ºC to 470ºC. The N-polar InN layers were grown at a significantly higher temperature of 550ºC on MBE-GaN grown on *c*-face SiC or the N-polar side of a freestanding GaN substrate. The free electron concentrations in the samples varied in the range from low-$10^{17}$ cm$^{-3}$ to low-$10^{18}$ cm$^{-3}$, and the carrier mobilities ranged from 2000 to 800 cm$^2$/Vs, as determined by single-field Hall measurements. The samples are labeled In(x) or N(x) for In- and N-polar growth, respectively, with x giving the growth regime In (In-droplets), s (stoichiometric) or N (N-rich). For more details on the growth procedures and electrical characterization, see Refs. [3, 4]. The samples used in the present study are described in Table 1. The positron experiments were performed at room temperature with a variable-energy positron beam. The Doppler broadening of the annihilation radiation was analyzed using the conventional *S* and *W* parameters. More information about the experimental setup can be found elsewhere [10].

The *S* and *W* parameters recorded in the InN layers are shown in Fig. 1. The sample In(N)-1 that produces the lowest *S* parameter of the present set of samples has an In vacancy content below the detection limit of positron annihilation spectroscopy [11], and is used as reference to which the data in Fig. 1 are normalized. The (*S*, *W*) points of the InN layers fall on the line connecting the characteristic parameters of the InN lattice and the In vacancy, taken from Ref. [9]. The position of the point on the line gives the concentration of the In vacancies—the closer the measured point is to the In vacancy, the more vacancies there are present. The concentration of the In vacancies ($V_{\text{In}}$) can be estimated from the *S* (or *W*) parameter data with the positron trapping model. When the cation vacancies are the only defects trapping positrons, their concentration can be determined from a simple formula (see Ref. [10]) by assuming a positron trapping coefficient of $\mu_V = 3 \times 10^{15}$ s$^{-1}$. The In vacancy concentrations in the measured InN layers range from below the room-temperature detection limit of about $1 \times 10^{16}$ cm$^{-3}$ to $7 \times 10^{16}$ cm$^{-3}$ as shown in Fig. 1 and Table 1.

The first observation to be made from the data is that the variations in $V_{In}$ concentration between the samples are surprisingly small regarding the differences in growth conditions. In particular, the growth temperature and stoichiometry, and the film thickness appear to have little impact on the vacancy concentration. It should also be noted that, given the large formation energy (> 3 eV) of $V_{In}$ in $n$-type InN [12], the equilibrium vacancy concentrations at the growth temperatures are expected to be much lower than observed (< $10^{13}$ cm$^{-3}$), suggesting that the incorporation of In vacancies during growth is dominated by other mechanisms than thermal formation. The second observation is that in both N-rich and In-rich growth of In-polar layers, the use of MOCVD-GaN as a growth template appears to result in a higher vacancy content than the use of an MBE-GaN buffer layer, in good agreement with results of optimizing the MBE-GaN buffer for InN growth presented in Ref. [3]. Looking more closely at the data, in the case of N-polar material, however, the growth stoichiometry does appear to have some effect on vacancy formation; the $V_{In}$ concentrations are lower in the samples grown in stoichiometric conditions, while In-rich growth seems to promote the formation of vacancy clusters: the data point of the sample N(In) appears to be off the line connecting the InN lattice and In-vacancy in Fig. 1. Interestingly, also in MOCVD growth of InN [8], vacancy clusters appear in the In-droplet regime. Another observable difference between the N- and In-polar samples (both grown on MBE-GaN) is that the vacancy concentrations are generally slightly larger in the N-polar films.

Considering the role of impurity atoms, there is no clear link between the predominant unintentionally incorporated oxygen and hydrogen impurities and $V_{In}$ formation, either. On the one hand, the O concentrations were found by secondary ion mass spectrometry (SIMS) to be about an order of magnitude larger in the N-polar films as compared to the In-polar layers on MBE-GaN. Moreover, comparing growth on MBE-GaN versus growth on MOCVD-GaN, previous work [3] has shown very high (~ $10^{20}$ cm$^{-3}$) peaks in the O and H concentrations near the film–substrate interface in direct growth on MOCVD-GaN—not present with the use of the MBE-GaN buffer layer—compared to the average impurity levels deep in the InN layers ([O] ~ $10^{17}$ cm$^{-3}$, [H] ~ $10^{18}$ cm$^{-3}$). On the other hand, regarding In-face InN, the impurity levels can be a few orders of magnitude higher when grown in N-rich conditions than in growth in the In-droplet regime [13], and yet based on the present set of samples, the $V_{In}$ concentrations are similar in both cases.

We also estimated the threading dislocation densities in the In-polar samples based on x-ray diffraction rocking curves [14]. The screw- and edge-type dislocation densities were determined to be in the ranges of 1–2 × $10^9$ cm$^{-2}$ and 1–5 × $10^{10}$ cm$^{-2}$, respectively, typical of high-quality InN [13, 14]. However, no dependence between the vacancy concentration and the dislocation densities could be established, suggesting that perhaps other structural defects, such as, e.g., stacking faults could be governing the $V_{In}$ formation. In fact, in GaN high densities of basal plane stacking faults and high $V_{Ga}$ concentrations have been observed to go hand in hand [15]. Furthermore, as in earlier work a link between the In vacancies and carrier mobility has been suggested [7], we show the relation of carrier mobility and In vacancy concentration in the samples where the electrical characteristics were measured in Fig. 2. In the present samples, we see no correlation between carrier mobility and [$V_{In}$] either, supporting instead the interpretation that extended defects are the most important scattering centers limiting the mobility in as-grown nominally undoped state-of-the-art MBE-InN [13].

The difference between the above observations in In-polar MBE-InN and those in Ga-polar MBE-GaN [16] is dramatic: the more N-rich the growth, the more Ga vacancies are formed, while no such effect on In vacancy formation is observed in InN. A similar difference has been observed in the case of MOCVD growth as well [8, 17]. A plausible explanation for the differences in cation-polar material can be found in the dominant formation mechanisms of the cation vacancies during growth in the two materials. In highly *n*-type material, i.e., Fermi level very close to the conduction band minimum (CBM), the calculated formation energy of the Ga vacancies in GaN is about 1.5 eV, while it is ~ 3 eV for the In vacancy in InN when the Fermi level is well above the CBM (corresponding to a free electron concentration > $10^{20}$ cm$^{-3}$) and ~ 5 eV when the Fermi level is close to the CBM (carrier concentration < $10^{19}$ cm$^{-3}$, as in our samples) [12, 18]. On the other hand, the growth temperature of GaN is 200–300ºC higher than that of InN in MBE growth and up to 500ºC higher in the case of MOCVD growth. The Ga vacancy concentrations in GaN samples with low dislocation densities are of the same order of magnitude that could be expected from the growth temperature and the calculated formation energies, given that the vacancies (which are mobile already at relatively low temperatures) are stabilized by impurities (such as O in GaN) or other defects relatively close to the growth temperature [15, 19, 20]. As the concentrations of the Ga and In vacancies are similar in samples in near-stoichiometric conditions, the formation of the In vacancies must be dictated by other factors than the thermal formation (and subsequent stabilization by, e.g., impurities) of an isolated In vacancy in an otherwise perfect lattice. On the other hand, assuming that the observed In vacancies were formed directly next to impurities or dislocations leads to an experimentally estimated formation energy of at most 0.7–0.9 eV [21]. The difference of more than 4 eV (i.e., the binding energy of the vacancy to the impurity or dislocation) to that calculated for the isolated In vacancy is much larger than that typically observed for point defect complexes in nitrides (1–2 eV) [18, 20]. Hence, this formation mechanism is not dominant either.

Another difference between MBE-InN and MBE-GaN can be found when comparing N-polar and In/Ga-polar growth. In the case of InN, the resulting vacancy concentrations are similar, while in GaN very high vacancy concentrations are observed in N-rich N-polar layers [16]. Interestingly, in Ga-rich N-polar MBE-GaN efficient clustering of vacancies was observed in that study, and a similar (although very small) effect is seen in the present results on In-rich N-polar MBE-InN. A possible explanation could be that the formation of N vacancies is strongly enhanced in Ga/In-rich N-polar growth, where a larger number of N sites are available to be left unfilled at the growth surface, but as the formation of In vacancies is much less probable than that of Ga vacancies, clustering is less efficient in InN.

Based on the above considerations, vacancy formation during the growth of PAMBE-InN might be dictated by surface adatom diffusion. As the observed vacancy concentrations are well above the thermal equilibrium concentrations, it seems plausible that the low growth temperature inhibits the lateral diffusion and thus the removal of vacancies at the surface during growth. Similarly to the case of GaN [22], the metallic surface adlayer during InN growth has a pronounced effect on the surface growth and the resulting film surface morphology [22–25]. In fact, the In surface adlayer coverage under In-rich conditions has been detemined to saturate at 1 ML for N-polar growth [24], resulting in a rougher film surface and higher impurity levels than in In-polar growth where an adlayer of up to 2.5 ML is observed [25]. Hence, the higher $V_{In}$ concentrations in the N-face than in the In-polar samples of the present set might be due to the lower surface adatom mobility (and hence less efficient In–N bond creation on the surface) rather than to the higher growth temperature of N-polar InN. Also, as the In adlayer coverage decreases with increasing N/In flux ratio, this could also explain the slightly higher vacancy content of the sample N(N) as compared to the other N-face samples.

To summarize, we have applied positron annihilation spectroscopy to study the effect of different growth conditions, such as growth polarity, temperature, and stoichiometry, film thickness, and substrate material on vacancy formation in InN grown by PA-MBE. All the InN layers studied in this work were found to contain In vacancies in the low- to mid-$10^{16}$ cm$^{-3}$ range. We conclude that the growth conditions, impurity concentrations and dislocation densities have little effect on the In vacancy concentrations in the films, but instead that the use of an optimized MBE-GaN buffer layer correlates with lower In vacancy concentration and higher carrier mobility. The results imply that in high quality InN, the electron mobility is likely limited not by ionized point defect scattering, but rather by threading dislocations. Finally, we suggest that the limited diffusion of surface adatoms may be the cause of the observation of In vacancy concentrations that are significantly higher than the estimated equilibrium concentrations during growth.

Acknowledgements:

This work has been partially funded by the MIDE programme and the Academy of Finland. Work at UCSB was supported by AFOSR (Dr. Kitt Reinhardt, Program Manager) and made use of the MRSEC Central Facilities supported by the National Science Foundation. The authors gratefully acknowledge Evans Analytical Group for SIMS data.

References:

[1] H. Lu, W. J. Schaff, J. Hwang, H. Wu, W. Yeo, A. Pharkya, and L. Eastman, Appl. Phys. Lett. **77**, 2548 (2000).

[2] Y. Saito, N. Teraguchi, A. Suzuki, T. Araki, and Y. Nanishi, Jpn. J. Appl. Phys., Part 2 **40**, L91 (2001).

[3] C. S. Gallinat, G. Koblmüller, J. S. Brown, S. Bernardis, J. S. Speck, G. D. Chern, E. D. Readinger, H. Shen, and M. Wraback, Appl. Phys. Lett. **89**, 032109 (2006).

[4] G. Koblmüller, C. S. Gallinat, S. Bernardis, J. S. Speck, G. D. Chern, E. D. Readinger, H. Shen, and M. Wraback, Appl. Phys. Lett. **89**, 071902 (2006).

[5] O. Briot, B. Maleyre, and S. Ruffenach, Appl. Phys. Lett. **83**, 2919 (2003).

[6] S. Suihkonen, J. Sormunen, V.-T. Rangel-Kuoppa, H. Koskenvaara, and M. Sopanen, J. Crystal Growth **291**, 8 (2006).

[7] J. Oila, A. Kemppinen, A. Laakso, K. Saarinen, W. Egger, L. Liszkay, P. Sperr, H. Lu, and W. J. Schaff, Appl. Phys. Lett. **84**, 1486 (2004).

[8] A. Pelli, K. Saarinen, F. Tuomisto, S. Ruffenach, and O. Briot, Appl. Phys. Lett. **89**, 011911 (2006).

[9] F. Tuomisto, A. Pelli, K. M. Yu, W. Walukiewicz, and W. J. Schaff, Phys. Rev. B **75**, 193201 (2007).

[10] K. Saarinen, P. Hautojärvi, and C. Corbel, in: Identification of Defects in Semiconductors, edited by M. Stavola, Semiconductors and Semimetals, Vol. 51A (Academic Press, New York, 1998) p. 209.

[11] F. Reurings, F. Tuomisto, C. S. Gallinat, G. Koblmüller, and J. S. Speck, Phys. Status Solidi (c) **6**, S401 (2009).

[12] C. Stampfl, C. G. Van de Walle, D. Vogel, P. Krüger, and J. Pollmann, Phys. Rev. B **61**, R7846 (2000).

[13] C. S. Gallinat, G. Koblmüller, and J. S. Speck, Appl. Phys. Lett. **95**, 022103 (2009).

[14] C. S. Gallinat, G. Koblmüller, Feng Wu, and J. S. Speck, J. Appl. Phys. **107**, 053517 (2010).

[15] F. Tuomisto, T. Paskova, R. Kröger, S. Figge, D. Hommel, B. Monemar, and R. Kersting, Appl. Phys. Lett. **90**, 121915 (2007).

[16] M. Rummukainen, J. Oila, A. Laakso, K. Saarinen, A. J. Ptak, and T. H. Myers, Appl. Phys. Lett. **84**, 4887 (2004).

[17] K. Saarinen, P. Seppälä, J. Oila, P. Hautojärvi, C. Corbel, O. Briot, and R. L. Aulombard, Appl. Phys. Lett. **73**, 3253 (1998).

[18] T. Mattila and R. M. Nieminen, Phys. Rev. B **55**, 9571 (1997).

[19] F. Tuomisto, K. Saarinen, B. Lucznik, I. Grzegory, H. Teisseyre, T. Suski, S. Porowski, P. R. Hageman, and J. Likonen, Appl. Phys. Lett. **86**, 031915 (2005).

[20] F. Tuomisto, K. Saarinen, T. Paskova, B. Monemar, M. Bockowski, and T. Suski, J. Appl. Phys. **99**, 066105 (2006).

[21] The equilibrium concentration is given by $c_V = N_{sites}\exp[(TS - E_f)/k_BT]$, where the entropy term $S$ is $\sim 5k_B$ (within reasonable limits, the calculated $c_V$ is not very sensitive to the magnitude of $S$), $E_f$ is the formation energy, and $N_{sites}$ is the atomic concentration of In for isolated vacancies and reduced to sites next to impurities (for maximum amount [O] = 1 × $10^{20}$ cm$^{-3}$, $N_{sites}$ = 4 × $10^{20}$ cm$^{-3}$). For dislocations we used the maximum observed density of 5 × $10^{10}$ cm$^{-2}$, resulting in a number of possible sites of $N_{sites}$ = 2 × $10^{20}$ cm$^{-3}$, where it is assumed that dislocations are able to bind vacancies at most two lattice constants away.

[22] G. Koblmüller, J. Brown, R. Aberbeck, H. Reichert, P. Pongratz, and J. S. Speck, Jpn. J. Appl. Phys., Part I **44**, L906 (2005).

[23] J. Neugebauer, T. K. Zywietz, M. Scheffler, J. E. Northrup, H. Chen, and R. M. Feenstra, Phys. Rev. Lett. **90**, 056101 (2003).

[24] G. Koblmüller, C. S. Gallinat, and J. S. Speck, J. Appl. Phys. **101**, 083516 (2007).

[25] C. S. Gallinat, G. Koblmüller, J. S. Brown, and J. S. Speck, J. Appl. Phys. **102**, 064907 (2007).

Table 1. Description and In vacancy concentrations of the PA-MBE InN samples. The carrier concentrations $n$ and mobilities $\mu$ were determined by Hall measurements.

| Sample | Growth $T$ (°C) | Thickness (nm) | Template | $[V_{In}]$ ($10^{16}$ cm$^{-3}$) | $n$ ($10^{18}$ cm$^{-3}$) | $\mu$ (cm$^2$V$^{-1}$s$^{-1}$) |
|---|---|---|---|---|---|---|
| In(In)-1 | 430 | 980 | MBE-GaN | 2 | 1.92 | 1336 |
| In(In)-2 | 450 | 250 | MOCVD-GaN | 6 | n/a | n/a |
| In(In)-3 | 450 | 750 | MOCVD-GaN | 3 | n/a | n/a |
| In(In)-4 | 470 | 1000 | MBE-GaN | 2 | 0.22 | 2176 |
| In(N)-1 | 430 | 830 | MBE-GaN | < 1 | 0.20 | 1224 |
| In(N)-2 | 450 | 300 | MOCVD-GaN | 5 | n/a | n/a |
| In(N)-3 | 450 | 1400 | MOCVD-GaN | 5 | 1.20 | 1513 |
| N(In) | 550 | 1000 | MBE-GaN / N-GaN | 5 | n/a | n/a |
| N(s)-1 | 550 | 500 | MBE-GaN / $c$-SiC | 3 | 0.9 | 1180 |
| N(s)-2 | 550 | 2000 | MBE-GaN / $c$-SiC | 4 | 0.48 | 1420 |
| N(N) | 550 | 1000 | MBE-GaN / $c$-SiC | 7 | 1.8 | 819 |

Figure captions:

Fig. 1. The average *S* and *W* parameters measured in the PA-MBE InN layers (normalized to the InN lattice values).

Fig. 2. The carrier mobility as a function of In vacancy content in the InN layers.

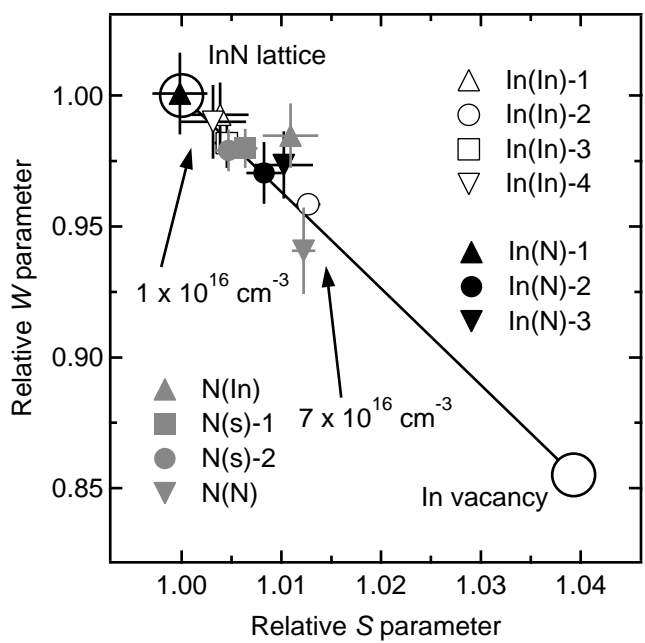

Fig. 1. Reurings et al.

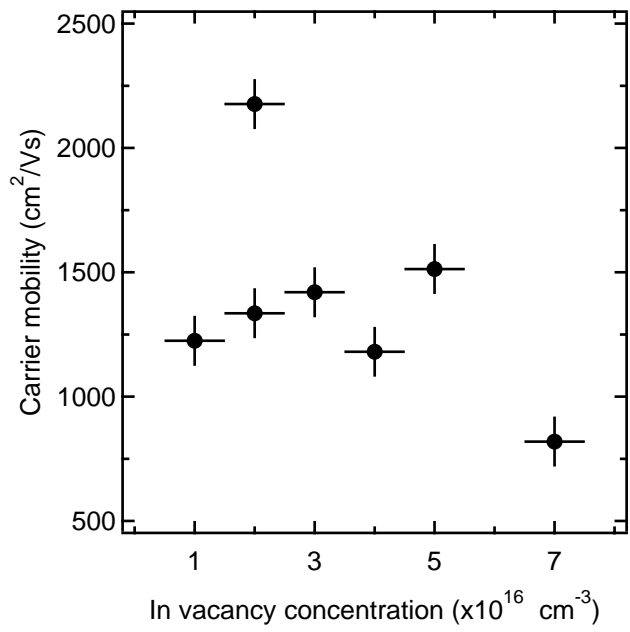

Fig. 2: Reurings et al.

15